\documentclass{revtex4-2}
\usepackage[utf8]{inputenc}
\usepackage{amsbsy}
\usepackage{amsopn}
\usepackage{amstext}
\usepackage{amsmath,amsthm,amsfonts,amssymb}
\usepackage[mathcal]{eucal}
\usepackage{mathrsfs}
\usepackage{braket}
\usepackage[english]{babel}
\usepackage{color}
\usepackage{esint}
\usepackage{graphicx}
\usepackage{float}
\usepackage{units}
\usepackage{blindtext}
\usepackage{hyperref}
\usepackage{textcomp}
\usepackage{orcidlink}

\DeclareGraphicsExtensions{.png,PNG,.pdf,.PDF}
\usepackage{hyperref}
\usepackage{slashed}
\newcommand{\ie}{\begin{equation}}
\newcommand{\fe}{\end{equation}}
\newcommand{\se}{\begin{eqnarray}}
\newcommand{\ff}{\end{eqnarray}}
\begin{document}

\title{Comment on ``Exact massless spinor quasibound states of Schwarzschild black hole''}
\author{R. R. S. Oliveira\,\orcidlink{0000-0002-6346-0720}}
\email{rubensrso@fisica.ufc.br}
\affiliation{Departamento de F\'isica, Universidade Federal da Para\'iba, Caixa Postal 5008, 58051-900, Jo\~ao Pessoa, PB, Brazil}


\date{\today}

\begin{abstract}

In this comment, we point out a series of errors made by Senjaya in your paper (2024) \cite{Senjaya}. In particular, these errors involved the Dirac equation in the 3+1 dimensional Schwarzschild spacetime, tetrad field, curved gamma matrices, and the time-independent Dirac equation written in terms of the inner/scalar product between the gamma matrices and the orbital angular momentum (i.e., $\vec{\gamma}\cdot\vec{L}$). Besides, the strange/peculiar thing about all this is that even Senjaya \cite{Senjaya} citing Collas and Klein \cite{Collas}, where the (mathematical) formalism is correct, everything indicates that he ``ignored'' or ``forgot'' to use such formalism in your paper. Therefore, using Ref. \cite{Collas}, we show here the correct form of the errors made by Senjaya \cite{Senjaya}.

\end{abstract}

\maketitle

\section{Introduction}

In a recent paper published in Physics Letters B, Senjaya \cite{Senjaya} investigated the behavior of massless spin $\frac{1}{2}$ field via the Dirac equation in a curved static spherically symmetric Schwarzschild spacetime. The author made a detailed derivation of the novel exact massive and massless scalar quasibound state in the Schwarzschild black hole background. After decoupling the Dirac equation, the author successfully solved both the angular and radial parts, respectively in terms of spin weighted Spherical Harmonics and Confluent Heun functions. With the exact radial wave solution in hand, and applying a polynomial condition, the author obtained the quantized energy levels expression. Besides, the author found that the massless spinor field around the Schwarzschild black hole has complex valued energy levels, in contrast to the purely imaginary for a massless boson around the same black hole. Finally, the author obtained the Hawking radiation distribution function (derived via the Damour-Ruffini method) and the Hawking temperature. In particular, this paper is well written and somehow seems to have some interesting results.

However, analyzing in detail a textbook cited/used by Senjaya \cite{Senjaya} about the Dirac equation in curved spacetimes or General Relativity (i.e., the curved Dirac equation), whose authors are Collas and Klein \cite{Collas}, we verified a series of errors made by Senjaya \cite{Senjaya}. In particular, these errors involved the Dirac equation in the 3+1 dimensional Schwarzschild spacetime, tetrad field, curved gamma matrices, and the time-independent Dirac equation written in terms of the inner/scalar product between the gamma matrices and the orbital angular momentum (i.e., $\vec{\gamma}\cdot\vec{L}$). So, the strange/peculiar thing about all this is that even Senjaya \cite{Senjaya} citing \cite{Collas}, where the (mathematical) formalism is correct, everything indicates that he ``ignored'' or ``forgot'' to use such formalism in your paper. Therefore, the goal of the present comment is to show the correct form of the errors made by Senjaya \cite{Senjaya}.


\section{Corrections of the Senjaya's paper}

According to Senjaya \cite{Senjaya}, the Dirac equation in curved spacetime is given by
\begin{equation}\label{1}
i\gamma^\mu\left(\partial_\mu-\frac{mc}{\hbar }\right)\psi=0,
\end{equation}
where $\gamma^\mu=\varepsilon^\mu_\alpha \gamma^\alpha_D$ are the curved gamma matrices (Dirac matrices in curved spacetime), $\varepsilon^\mu_\alpha$ is the tetrad field or simply tetrads (a matrix that connects both curved and flat spacetimes), $\gamma^\alpha_D$ are the flat gamma matrices (Dirac matrices in the Minkowsky spacetime), $m$ is the spinor mass and $c$ is the speed of light, respectively.

However, according to Ref. \cite{Collas}, the above equation is wrong/incomplete. For example, according to \cite{Collas} (Chapter 3: The Spinorial Covariant Derivative; Eqs. (3.31)-(3.32)) the Dirac equation in curved spacetime is given by ($\hbar=c=1$)
\begin{equation}\label{2}
i\bar{\gamma}^\mu D_\mu\psi-m\psi=0,
\end{equation}
where $D_\mu=\partial_\mu+\Gamma_\mu$ is the spinor covariant derivative (or simply the covariant derivative), $\Gamma_\mu=\frac{\varepsilon}{2}\omega_{AB\mu}\Sigma^{AB}$ ($\varepsilon=\pm 1$) is the spinor affine connection, $\omega_{AB\mu}$ is the spin connection, and $\bar{\gamma}^\mu=\bar{\gamma}^\mu(x)=e^{\ \mu}_A (x)\gamma^A$ are curved gamma matrices, being $e^{\ \mu}_A (x)$ the tetrads and $\gamma^A$ the flat gamma matrices. Therefore, we clearly see that Senjaya \cite{Senjaya} ``ignored'' or ``forgot'' to use the covariant derivative of \cite{Collas} in the Dirac equation of your paper. Furthermore, even if Senjaya \cite{Senjaya} had used such a covariant derivative, i.e.: $i\gamma^\mu(D_\mu-mc/\hbar)\psi=0$, the Dirac equation would still be wrong since the gamma matrices $\gamma^\mu$ (or better, $i\gamma^\mu$) should only act (multiply) in the covariant derivative $D_\mu$ and not in the term constant $mc/\hbar$ (such as it appears in \eqref{2}). For more (valuable) information about the correct form for the curved Dirac equation (such as in \eqref{2}), we recommend Refs. \cite{Lawrie,oliveira1,oliveira2,oliveira3,oliveira4,Panahi,Yamamoto,Alsing,Pollock,Chen,Ebihara,Chernodub1,Cao,Oliveira,Zhang,Chen2,Ayala,Sadooghi,Hammad,Tabatabaee,Mehr,Nyambuya}. 

Furthermore, according to \cite{Collas}, the tetrads of Senjaya \cite{Senjaya} are wrong (actually one of its components). For example, according to \cite{Collas} (Chapter 4: Examples in (3+1) GR; Eqs. (4.3)-(4.6)), the tetrads for the Schwarzschild spacetime are given by
\begin{equation}\label{3}
e^{\ \mu}_A (x)=diag(e^{\ t}_0 (x),e^{\ r}_1 (x),e^{\ \theta}_2 (x),e^{\ \phi}_3 (x))=diag\left(\frac{1}{\sqrt{f}},\sqrt{f},\frac{1}{r},\frac{1}{r\sin\theta}\right), \ \ f=f(r)=1-\frac{r_s}{r},
\end{equation}
where $r_s=2M$ is the Schwarzschild radius.

That is, Senjaya \cite{Senjaya} wrote the component $e^{\ \theta}_2 (x)$ wrongly, given by $e^{\ \theta}_2 (x)=\frac{1}{\sin\theta}$. Therefore, the correct form for the matrix $\gamma^\mu$ in your paper should be
\begin{equation}\label{4}
\begin{pmatrix}
\gamma^0 \\
\gamma^1 \\
\gamma^2 \\
\gamma^3
\end{pmatrix} = 
\begin{pmatrix}
\frac{1}{\sqrt{f}} & 0 & 0 & 0 \\
0 & \sqrt{f} & 0 & 0 \\
0 & 0 & \frac{1}{r} & 0  \\
0 & 0 & 0 & \frac{1}{r\sin\theta} 
\end{pmatrix}\begin{pmatrix}
\gamma^0_D \\
\gamma^1_D  \\
\gamma^2_D  \\
\gamma^3_D 
\end{pmatrix} .
\end{equation}

So, using Eq. \eqref{1}, Senjaya \cite{Senjaya} obtained the Dirac equation in the static spherically symmetric spacetime, written as follows
\begin{equation}\label{5}
\left[\frac{i}{\sqrt{f}}\gamma^0_D \partial_0+i\sqrt{f}\gamma^1_D \partial_1+\frac{i}{\sin\theta}\gamma^2_D\partial_2+\frac{i}{r\sin\theta}\gamma^3_D \partial_3-\frac{mc}{\hbar}\right]\psi=0.
\end{equation}

However, the ``correct’’ (quotes because $\Gamma_\mu$ was ignored) according to \eqref{4}, it would be
\begin{equation}\label{6}
\left[\frac{i}{\sqrt{f}}\gamma^0_D \partial_0+i\sqrt{f}\gamma^1_D \partial_1+\frac{i}{r}\gamma^2_D\partial_2+\frac{i}{r\sin\theta}\gamma^3_D \partial_3-\frac{mc}{\hbar}\right]\psi=0.
\end{equation}

In addition, using the following definitions \cite{Senjaya}
\begin{equation}\label{7}
\frac{i}{\sin\theta}\gamma^2_D\partial_2+\frac{i}{r\sin\theta}\gamma^3_D \partial_3=\frac{1}{\hbar}\vec{\gamma}\cdot(\hat{r}\times\vec{L}),
\end{equation}
\begin{equation}\label{8}
(\vec{\gamma}\cdot\vec{A})(\vec{\gamma}\cdot\vec{B})=\gamma^0\vec{A}\cdot\vec{B}+\frac{i}{\hbar}\vec{\gamma}\cdot(\vec{A}\times\vec{B}),
\end{equation}
implies that \cite{Senjaya}
\begin{equation}\label{9}
\frac{i}{\sin\theta}\gamma^2_D\partial_2+\frac{i}{r\sin\theta}\gamma^3_D \partial_3=-\frac{i}{\hbar r}\gamma^1(2\vec{\gamma}\cdot\vec{L}).
\end{equation}

Consequently, Eq. \eqref{5} becomes \cite{Senjaya}
\begin{equation}\label{10}
\left[\frac{1}{\sqrt{f}}\gamma^0_D\frac{E}{\hbar c}+\gamma^1_D\left(i\sqrt{f}\partial_1-\frac{i}{\hbar r}(2\vec{\gamma}\cdot\vec{L})\right)\right]\Psi=0,
\end{equation}
where $\psi=e^{-\frac{iE}{\hbar c}ct}\Psi(r,\theta,\phi)$ and $m=0$. In particular, Senjaya \cite{Senjaya} made a little mistake here because used $\psi$ instead of $\Psi$ as the solution to Eq. \eqref{10}.

So, it is with Eq. \eqref{10} that Senjaya \cite{Senjaya} develops your entire paper. However, this equation is wrong since the contribution of $\Gamma_\mu$ was completely ignored. Consequently, the results of Senjaya's paper \cite{Senjaya} were negatively impacted (i.e., are also wrong).

Now, let us correct Eq. \eqref{10} through the contribution of $\Gamma_\mu$, that is, let us obtain the correct equation for Senjaya's paper \cite{Senjaya}. So, according to \cite{Collas} (Chapter 4: Examples in (3+1) GR; Eqs. (4.7)-(4.14)), the nonvanishing spin connections for the Schwarzschild spacetime are given by
\begin{equation}\label{11}
\omega_{10t}=\frac{r_s}{2r^2},
\end{equation}
\begin{equation}\label{12}
\omega_{21\theta}=\sqrt{f},
\end{equation}
\begin{equation}\label{13}
\omega_{31\phi}=\sin\theta\sqrt{f},
\end{equation}
\begin{equation}\label{14}
\omega_{32\phi}=\cos\theta,
\end{equation}
while the nonvanishing spinor affine connections (obtained through the Fock-Ivanenko coefficients $\Gamma_C=e_C^{\ \mu} \Gamma_\mu$) are given by
\begin{equation}\label{15}
\Gamma_0=\frac{r_s}{4r^2}\gamma^0_D\gamma^1_D,
\end{equation}
\begin{equation}\label{17}
\Gamma_2=\frac{1}{2}\sqrt{f}\gamma^1_D\gamma^2_D,
\end{equation}
\begin{equation}\label{18}
\Gamma_3=\frac{\sin\theta}{2}\sqrt{f}\gamma^1_D\gamma^3_D+\frac{\cos\theta}{2}\gamma^2_D\gamma^3_D.
\end{equation}

So, doing $\partial_\mu\to \partial_\mu+\Gamma_\mu$ in Eq. \eqref{6}, we obtain
\begin{equation}\label{19}
\left[\frac{\gamma^0_D}{\sqrt{f}}\left(i\partial_0+\frac{r_s}{4r^2}\gamma^0_D\gamma^1_D\right)+i\sqrt{f}\gamma^1_D \partial_1+\frac{i\gamma^2_D}{r}\left(\partial_2+\frac{1}{2}\sqrt{f}\gamma^1_D\gamma^2_D\right)+\frac{i\gamma^3_D}{r\sin\theta}\left(\partial_3+\frac{\sin\theta}{2}\sqrt{f}\gamma^1_D\gamma^3_D+\frac{\cos\theta}{2}\gamma^2_D\gamma^3_D\right)-\frac{mc}{\hbar}\right]\psi=0.
\end{equation}

Therefore, using the relation \eqref{9} (with $e^{\ \theta}_2 (x)=1/r$) into \eqref{19} with $\psi=e^{-\frac{iE}{\hbar c}ct}\Psi(r,\theta,\phi)$ and $m=0$, we obtain the following corrected equation for the Senjaya’s paper \cite{Senjaya}
\begin{equation}\label{20}
\left[\frac{\gamma^0_D}{\sqrt{f}}\left(\frac{E}{\hbar c}+\frac{r_s}{4r^2}\gamma^0_D\gamma^1_D\right)+\gamma^1_D\left(i\sqrt{f}\partial_1-\frac{i}{\hbar r}(2\vec{\gamma}\cdot\vec{L})\right)+\frac{i\gamma^2_D}{r}\left(\frac{1}{2}\sqrt{f}\gamma^1_D\gamma^2_D\right)+\frac{i\gamma^3_D}{r\sin\theta}\left(\frac{\sin\theta}{2}\sqrt{f}\gamma^1_D\gamma^3_D+\frac{\cos\theta}{2}\gamma^2_D\gamma^3_D\right)\right]\Psi=0.
\end{equation}


\section{Final remarks}

In this comment, we point out a series of errors made by Senjaya \cite{Senjaya} in your paper. In particular, these errors involved the Dirac equation in the 3+1 dimensional Schwarzschild spacetime, tetrad field, curved gamma matrices, and the time-independent Dirac equation written in terms of the inner/scalar product between the gamma matrices and the orbital angular momentum (i.e., $\vec{\gamma}\cdot\vec{L}$). Besides, the strange/peculiar thing about all this is that even Senjaya \cite{Senjaya} citing \cite{Collas}, where the (mathematical) formalism is correct, everything indicates that he ``ignored'' or ``forgot'' to use such formalism in your paper. Therefore, using Ref. \cite{Collas}, we show here the correct form of the errors made by Senjaya \cite{Senjaya}.
 
\section*{Acknowledgments}

\hspace{0.5cm}The author would like to thank the Conselho Nacional de Desenvolvimento Cient\'{\i}fico e Tecnol\'{o}gico (CNPq) for financial support through the postdoc grant No. 175392/2023-4, and also to the Department of Physics at the Universidade Federal da Para\'{i}ba (UFPB) for hospitality and support.

\section*{Data availability statement}

\hspace{0.5cm} This manuscript has no associated data or the data will not be deposited. [Author’ comment: There is no data associated with this manuscript or no data has been used to prepare it.]

\end{document}